# Observation of REBCO delamination in the resistive insulation nested coils

**Jun Lu\*, Iain Dixon, Kwangmin Kim, Yan Xin, and Hongyu Bai**

Magnetic Science and Technology, National High Magnetic Field Laboratory, Tallahassee, Florida, USA

*E-mail: junlu@magnet.fsu.edu

**Abstract.** The REBCO coated conductor has the potential to be widely used in ultrahigh field magnets. It is well known, however, that it is not mechanically strong against delamination in the direction normal to its surface due to its intrinsic layered structure. Therefore, conductor delamination is one of the major design challenges for REBCO magnet coils. As a part of the development of the 40 T all-superconducting magnet at the National High Magnetic Field Laboratory, USA (NHMFL), a dry-wound resistive-insulation-nested-coils (RINC) was designed to reach 25.8 T. It used surface-treated stainless-steel tape as a co-wind to control the turn-to-turn contact resistance, and was fabricated and tested in a liquid helium bath. During the test, two of the double pancake modules exhibited resistive transitions at a current significantly lower than the designed value. The postmortem inspection of the REBCO conductor of these modules by reel-to-reel magnetization at 77 K found sections of very low critical current. Further investigations of one section by chemical etching, visual inspection, and electron microscopy revealed that conductor of this section was delaminated. We present the detailed findings of these postmortem characterizations. The implication of this type of delamination for future magnet designs will be discussed.

## 1. Introduction

REBCO coated conductor, as a layered composite tape, is prone to delamination which is a major challenge for REBCO magnet design. The severe impact of the delamination has been demonstrated in epoxy impregnated coils [1], which has stimulated a significant amount of research on characterization and mitigation of delamination by many research groups worldwide [2]-[7]. To minimize the probability of delamination by thermal stress, paraffin impregnated [8], and dry-wound coils [9] were operated successfully up to 45 T. On the other hand, recent experimental results suggest that delamination can be introduced by electromagnetic stress alone in high magnetic fields [10]-[13]. This makes the situation much more complicated because the screening currents can generate tensile transverse stress in half of the conductor width even if the designed electromagnetic stress is compressive. It becomes difficult to avoid delamination unless the delamination strength of the REBCO is improved to become much greater than the electromagnetic stress. However, the delamination in dry-wound coils, presumably due to transverse electromagnetic stress, has not been reported. In this paper, we report our observation of a delaminated section in a dry-wound resistive-insulation-nested-coils (RINC) which degraded the performance of the magnet significantly. The implication to future magnet designs and the suggested mitigation method are discussed.

## 2. Experimental methods

The RINC magnet was designed as a prototype of the HTS insert coil for the 40 T project. The magnet design parameters are shown in Table 1. The double-pancakes (modules) were wound by using SuperPower SCS4050-AP 4 mm wide REBCO tapes co-wound with 25 μm thick 316 stainless-steel tapes as reinforcement. For contact resistivity control, REBCO was coated by solder by dipping it in a 240 °C molten eutectic PbSn solder pot. The surface of the stainless-steel co-wind tapes were slightly oxidized by heating in air at different temperatures to achieve two different levels of contact resistivity [14]. It should be noted that the pancakes were wound with REBCO facing the inner diameter of the winding, so the apparent transverse (direction normal to the broad face) electromagnetic stress on REBCO was compressive during the magnet charging. Figure 1 shows the mechanical design and a picture of the fabricated RINC magnet.

**Table 1.** RINC magnet design parameters.

| Parameter | Coil 1 | Coil 2 |
|---|---|---|
| Central field | 25.8 T | |
| Current | 315 A | |
| Inductance | 4.15 H | |
| Stored energy | 206 kJ | |
| Cold bore | 34 mm | |
| Cold mass | 98 kg | |
| Inner/outer radius | 20/40 mm | 50/75 mm |
| Number of DP | 12 | 18 |
| Number of turns /DPs | 155 | 193 |
| Contact resistivity | 1 mΩ-cm$^2$ | 5 mΩ-cm$^2$ |

The RINC magnet was tested in liquid nitrogen to verify the low resistances of crossovers between double-pancake modules. Then it was warmed up and cooled down to 4.2 K for the final test. At 210 A, which is much lower than the targeted operating current of 315 A, resistive voltages appeared in module 1 of coil 1 and module 5 of coil 2 as shown in the Figure 2 indicating the degradation of these two modules.

The degraded modules were disassembled from the RINC and unwound. The REBCO tapes used in these two modules were subsequently inspected by the remnant magnetization method at 77 K using a reel-to-reel TapeStar™ system at SuperPower Inc. This was followed by a more detailed inspection using a similar reel-to-reel magnetization inspection device developed at our laboratory. The degradations observed by magnetization were confirmed by using transport critical current ($I_c$) measurements at 77 K in self-field using a conventional four-probe method. In this test, pressure contacts were used for voltage taps which were 50 mm apart. The criterion to determine $I_c$ was 1 µV/cm.

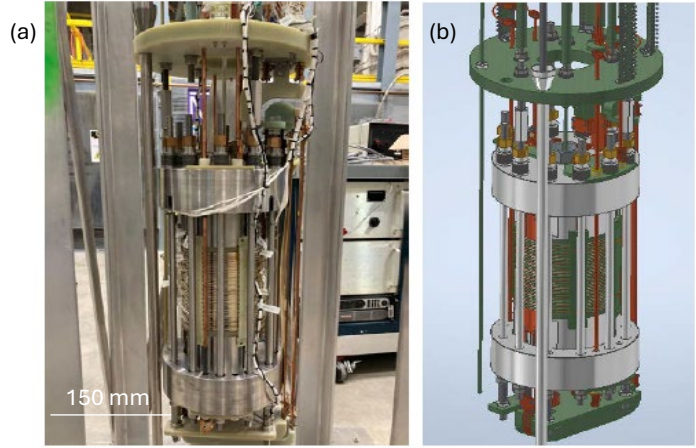

**Figure 1.** (a) picture of the RINC magnet after coil assembly. (b) the CAD drawing of the RINC.

To prepare for electron microscopy analysis, the solder coating was removed by a chemical solder stripper ( Hubbard-Hall Inc).  Then the copper layer on the REBCO sample was removed chemically by using APS-100 copper etchant (Transene CO Inc.) at room temperature. Cross-sectional TEM samples were prepared by focused ion beam (FIB) in a Thermal Scientific Helios G4 dual-beam field-emission scanning electron microscope (SEM). The TEM samples were subsequently studied by scanning transmission electron microscopy (STEM) using the probe-aberration-corrected JEOL JEM-ARM200cF at 200 kV. High-angle annular scanning transmission electron microscopy (HAADF-STEM) imaging and the annual-bright-field STEM (ABF-STEM) imaging were the two major imaging techniques used.

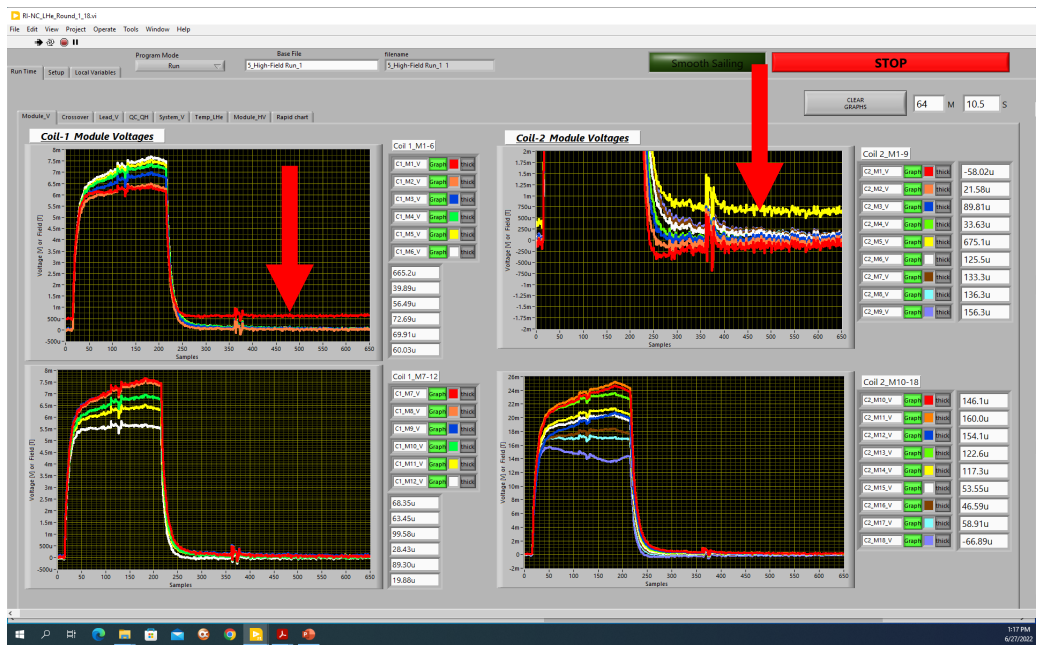

**Figure 2.** A screenshot of the data acquisition software during RINC coil test at 4.2 K and 210 A. The non-zero voltages of module 1 of coil 1 (the arrow on the left panel) and module 5 of coil 2 (the arrow on the right panel) indicate degradation of REBCO conductors in these two modules.

## 3. Results and discussions

The TapeStar™ data along the length of the REBCO used in the degraded module 1 of coil 1 are shown in Figure 3. A low $I_c$ location is clearly identified as indicated in the figure by the arrow. This was confirmed by a detailed magnetization measurement using our in-house magnetization test device. The section of about 328 mm that included the spot with very low $I_c$ was cut and separated into three pieces (A - C) and each was tested by the 77 K self-field transport measurement. The results of these three measurements are shown in Figure 4. As shown in the figure, sample B and sample C were severely degraded with considerably low $I_c$, in agreement with the TapeStar™ result.

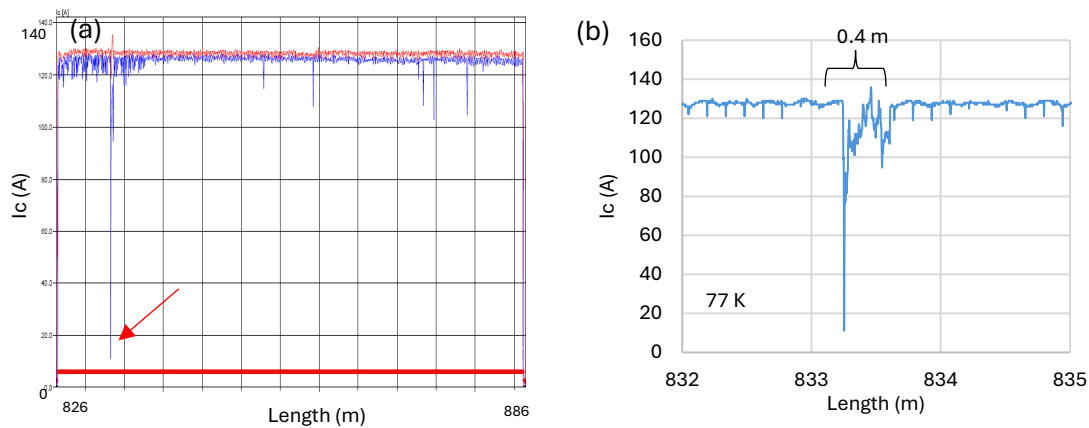

**Figure 3.** Reel-to-reel remnant magnetization of REBCO tape at 77 K for module 1 of coil 1. (a) the TapeStar™ data plot. (b) close-up plot near the dropout indicated by the arrow in (a).

None of the three samples showed any visible damages or defects on their solder coated surface. After the solder removal, the samples still did not show any surface damages or defects. Then the copper was etched to reduce the amount of FIB cutting needed to make cross-sectional TEM samples. As the copper was gradually etched away, however, defects became clearly visible in sample B and C. As shown in Figure 5, the surface of undegraded sample A is smooth consists of silver and defect free. Whereas sample B has a region of delamination near the edge, so the etching was terminated to prevent further chemical reaction to the REBCO layer. Sample C shows an interesting surface morphology with a string of delaminated bubbles on one side of the tape.

Cross-section TEM samples made by FIB on one of the bubbles indicated by the arrow in Figure 6. As shown in the cross-sectional SEM image in Figure 7, the bubble is a delaminated region. The delamination occurred within the REBCO layer with about ¾ of the total thickness separated from the remaining REBCO which is still attached to the substrate. TEM image is Figure 8 provides a high resolution view of the same region. These cross-sectional microscopy images provide strong evidence that the delamination occurred mostly within the REBCO layer near the REBCO/buffer interface.

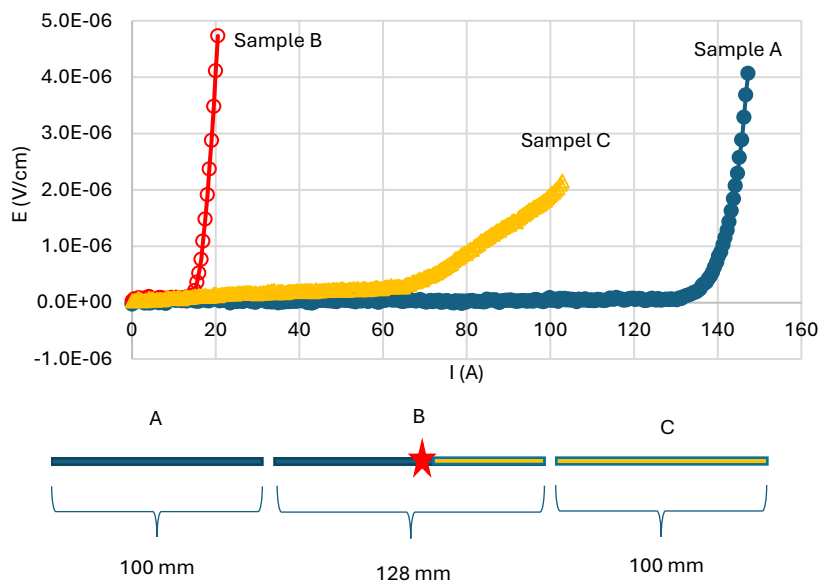

**Figure 4.** V-I curves measured at 77 K self-field for samples A - C. The red start in the schematic drawing corresponds to the sharp dip in magnetization signal shown in Figure 3.

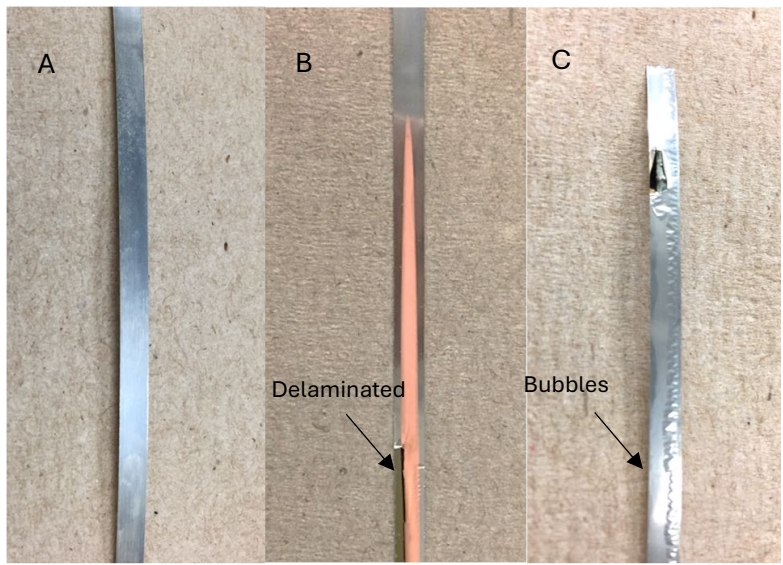

**Figure 5.** Sample A, B, C after etching by APS-100 copper etchant. Sample B was not completely etched because some part of the tape already delaminated and the REBCO layer was exposed as indicated by the arrow. Sample C showed a train of delaminated bubbles.

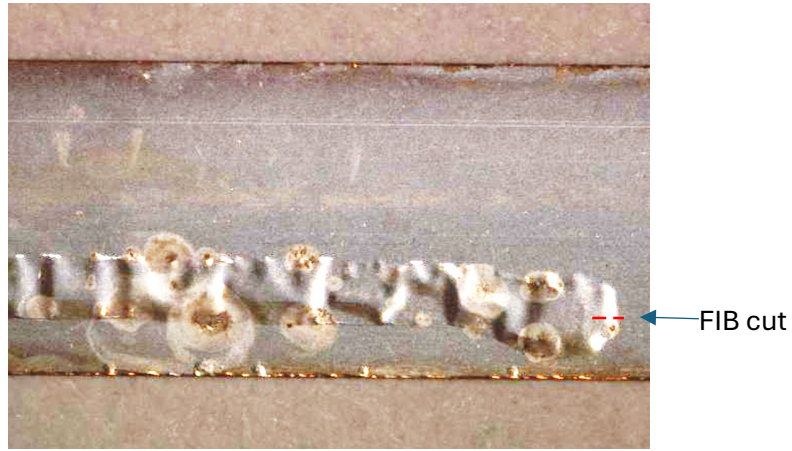

**Figure 6.** The bubbles seen in sample C. The arrow indicates the location where TEM samples were cut by FIB.

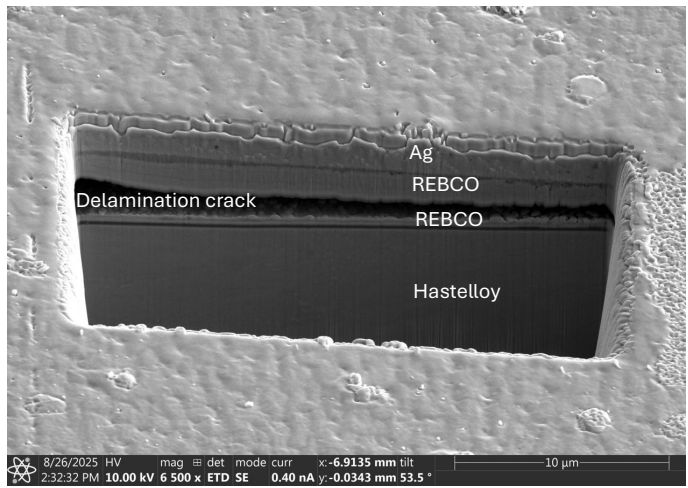

**Figure 7.** SEM image of a FIB cross-section of sample C. The delamination crack is seen within the REBCO layer.

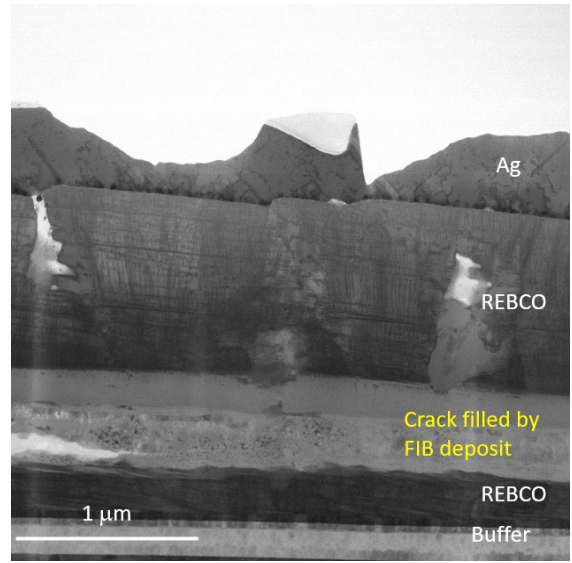

**Figure 8.** ABF-STEM image of the delaminated REBCO in sample C.

## 4. Discussions

The observation of the bubbles of delamination, which most likely occurred during the magnet charging at liquid helium temperature, substanciated the concern of delamination of REBCO in ultra-high field magnets. Although in this case the nominal transverse stress was compressive, i.e. the electromagnetic force push the REBCO layer towards the Hastelloy substrate, the ramping magnetic field created a screening current which resulted in tensile stress in one half of the tape width. This is supported by our recent experimental demonstration of delamination by screening currents [11]. In this proposed delamination mechanism, the modules near two ends of the solenoid magnet experience highest transverse tensile stress because of high field angles and high induced screening current, while the modules near the mid-plane will have much lower transverse tensile stress due to low screening current.

To mitigate the risk of such delaminations, it is critically important to improve the delamination strength of the REBCO conductor significantly. The challenge is also for the magnet designers to specify the minimum delamination strength, and for the quality control system to capture the sections of low delamination strength before they made their way into a magnet coil. In addition to improvement of REBCO, another mitigation method may also be effective. For instance, the fact that the delamination was not visualy observable until the removal of copper suggests that thicker copper layer might be helpful to mitigate the probabilty of delamination. This point is supported by recent report of delamination strength increase with copper thickness [15].

It is interesting to observe that the delamination bubbles were not visible until the copper layer was removed. Apparently the 20 μm copper and the adjacent turn in the coil provided constraint, and the delamination did not cause observable local plastic deformation to the copper layer. When the copper was removed the delaminated regions were under 1 – 2 μm of silver. Bubbles became clearly visible, suggesting severe plastic deformation of the silver layer during etching possibly by the hydrolic pressure from the gas that was trapped in the bubbles. Is it possible for gas (presumably helium) to permeate through the surrounding copper layer? To answer this question, future experiments are planned to identify the content in the delamination bubbles.

## 5. Summary

Significant $I_c$ degradation was found in the RINC magnet during its test in liquid helium. The TapeStar™ was used successfully to locate the degraded region. Transport $I_c$ test at 77 K confirmed the TapeStar™ results. A combination of chemical etching, SEM, and TEM investigation revealed that this $I_c$ degradation was due to delamination bubbles, which is visible after chemical removal of the copper layer. The cause for the delamination is attributed to electromagnetic stress introduced by screening current.

## Acknowledgments

We thank our colleagues at the NHMFL who worked on the RINC design, manufacturing, and testing: Denis Markiewicz, Peng Xu, Dylan Kolb-Bond, Kurt Cantrell, Scott Marshall, Lee Marks, Brent Jarvis, Justin Deterding, Erick Aroyo, and Mark Bird. We also thank David Greene for the development of magnetization test for the NHMFL and SuperPower for postmortem inspection by TapeStar™.

This work was performed at the National High Magnetic Field Laboratory, which is supported by National Science Foundation Cooperative Agreement No. DMR- 2128556, DMR-2131790, and the State of Florida.